\documentclass[12pt]{article}
\usepackage{amsmath,amssymb,amsthm}
\usepackage{graphicx,subfigure}
\usepackage[small]{caption}
\usepackage{changepage}
\usepackage{hyperref,cite,url,breakurl}
\usepackage{mathtools,xparse,MnSymbol}
\usepackage{multirow}
\usepackage{stackrel}
\begin{document}
\baselineskip 0.6cm

\def\bra#1{\langle #1 |}
\def\ket#1{| #1 \rangle}
\def\inner#1#2{\langle #1 | #2 \rangle}
\def\brac#1{\llangle #1 \|}
\def\ketc#1{\| #1 \rrangle}
\def\innerc#1#2{\llangle #1 \| #2 \rrangle}
\def\app#1#2{%
  \mathrel{%
    \setbox0=\hbox{$#1\sim$}%
    \setbox2=\hbox{%
      \rlap{\hbox{$#1\propto$}}%
      \lower1.1\ht0\box0%
    }%
    \raise0.25\ht2\box2%
  }%
}
\def\approxprop{\mathpalette\app\relax}
\DeclarePairedDelimiter{\norm}{\lVert}{\rVert}

\begin{titlepage}

\begin{flushright}
\end{flushright}

\vskip 1.2cm

\begin{center}
{\Large \bf The Interior of a Unitarily Evaporating Black Hole}

\vskip 0.7cm

{\large Yasunori Nomura}

\vskip 0.5cm

{\it Berkeley Center for Theoretical Physics, Department of Physics,\\
  University of California, Berkeley, CA 94720, USA}

\vskip 0.2cm

{\it Theoretical Physics Group, Lawrence Berkeley National Laboratory, 
 Berkeley, CA 94720, USA}

\vskip 0.2cm

{\it Kavli Institute for the Physics and Mathematics of the Universe 
 (WPI),\\
 UTIAS, The University of Tokyo, Kashiwa, Chiba 277-8583, Japan}

\vskip 0.8cm

\abstract{We study microscopic operators describing the experience of 
an observer falling into the horizon of a unitarily evaporating black 
hole. For a young black hole, these operators can be taken to act only 
on the degrees of freedom in the black hole region:\ the soft---or 
stretched horizon---modes as well as the semiclassical modes in the 
zone region. On the other hand, for an old black hole, the operators 
must also involve radiation emitted earlier; the difference between 
the two cases comes from statistics associated with the coarse-graining 
performed to obtain the effective theory of the interior. We find 
that the operators relevant for the interior theory can be defined 
globally as standard linear operators throughout the microstates, 
which obey the correct algebra up to corrections exponentially 
suppressed in the ratio of excitation energy to the Hawking 
temperature. We conjecture that the existence of such global 
operators is required for the emergence of the semiclassical picture. 
We also elucidate relation between the present construction and 
entanglement wedge reconstruction of the interior.}

\end{center}
\end{titlepage}

\section{Introduction}
\label{sec:intro}

Since the discovery of the thermodynamics of a black 
hole~\cite{Bekenstein:1973ur,Hawking:1974sw}, studying its 
physics has been driving our understanding of spacetime and 
gravity at the fundamental level.  In particular, addressing 
the apparent contradiction~\cite{Hawking:1976ra} between the 
thermodynamic aspects of a black hole and the principles of 
quantum mechanics has led to many important ideas, including 
black hole complementarity~\cite{Susskind:1993if}.  While we 
have found that the principles of quantum mechanics prevail 
in the end~\cite{tHooft:1990fkf,Maldacena:1997re}, paradoxes 
still remain.  One of the latest forms of these is the difficulty 
of reconciling unitary evolution of a black hole with the existence 
of its interior~\cite{Almheiri:2012rt}.

The purpose of this paper is to study how operators describing the 
interior of a unitarily evaporating black hole can be constructed at 
the microscopic level.  For a black hole in asymptotically flat spacetime 
(or a small black hole in AdS spacetime), we work in the framework of 
Refs.~\cite{Nomura:2018kia,Nomura:2019qps}.  This framework describes 
a black hole as a state in which the hard modes---the modes relevant 
for describing small objects around the black hole---are entangled in 
a generic manner with the soft modes---the degrees of freedom comprising 
the majority of the black hole.  This entanglement is generated by 
a strong, chaotic dynamics near the horizon.  We will also discuss 
an extension of the framework to a large AdS black hole.

A black hole forms when the system-specific properties, e.g.\ the details 
of the surface of a collapsing star, are strongly redshifted due to gravity, 
so that the system appears to be in a vacuum state at the semiclassical 
level.  This redshift makes the majority of the degrees of freedom unobservable 
for a long time.  These ``hidden'' degrees of freedom, associated with the 
black hole, have an exponentially large density of states~\cite{Bekenstein:1973ur} 
and are distributed mostly near the horizon.  While they are very soft 
when measured in the asymptotic region, their intrinsic dynamical scale 
is larger near the horizon due to gravitational blueshift, reaching the 
string scale at the stretched horizon~\cite{Susskind:1993if}.  The dynamics 
in this region is chaotic~\cite{Hayden:2007cs,Sekino:2008he} across 
all low energy species, giving generic entanglement between the hidden 
degrees of freedom (soft modes) and the others which can be used to 
describe small excitations around the black hole (hard modes).  In fact, 
this entanglement is the origin~\cite{Nomura:2018kia} of the thermality 
of the black hole~\cite{Hawking:1974sw}.

While the strong dynamics near the stretched horizon cannot be described 
by a low energy quantum field theory, we expect that it is unitary, as 
implied by the AdS/CFT correspondence~\cite{Maldacena:1997re}.  The fact 
that this dynamics lies outside the validity of the low energy theory 
allows for avoiding the conclusion of information loss~\cite{Hawking:1976ra} 
reached by semiclassical calculation.  The picture of the black hole 
interior can emerge through a coarse-graining of the degrees of freedom 
that cannot be physically resolved by an infalling observer, whose 
consistency with black hole's unitary evolution is ensured by a specific 
entanglement structure between the hard modes, soft modes, and early 
radiation~\cite{Nomura:2019qps}.

In this paper, we analyze how the operators in the coarse-grained, 
effective theory of the interior can be realized in the original 
microscopic theory.  We show that the construction of these operators 
is not unique.  In particular, we find that the operators can be 
written without involving an element outside the black hole region 
for a young black hole, a black hole that is not yet maximally entangled 
with the early radiation.  On the other hand, for an old black hole, 
radiation degrees of freedom must be involved, despite the fact that 
the degrees of freedom describing a falling object are not directly 
entangled with the early radiation.  We study how this happens both in 
the cases of a black hole in asymptotically flat spacetime and a large 
black hole in AdS spacetime.  This analysis elucidates which features 
of the analyses using a large AdS black hole may apply to a black hole 
in asymptotically flat spacetime.

The construction described in this paper addresses all the issues 
raised regarding the interior of a unitarily evaporating black 
hole~\cite{Almheiri:2012rt,Almheiri:2013hfa,Marolf:2013dba,%
Bousso:2013ifa,Marolf:2015dia}.  In particular, we find that all 
the operators relevant for the effective theory of the interior 
can be represented as standard linear quantum operators at the 
microscopic level, which obey the correct algebra up to corrections 
exponentially suppressed in the ratio of excitation energy to the 
Hawking temperature, which is typically a very large number.  We 
also argue that the construction preserves spacetime locality at 
the semiclassical level.

The organization of this paper is as follows.  In 
Section~\ref{sec:framework}, we describe the framework of 
Refs.~\cite{Nomura:2018kia,Nomura:2019qps}, highlighting features 
relevant for our discussion.  In Section~\ref{sec:interior}, we present 
our analysis of interior operators for a black hole in asymptotically 
flat spacetime (or a small black hole in AdS spacetime).  We construct 
operators that play the role of annihilation and creation operators 
in the effective theory of the interior erected at a given time in 
a distant description.  We find that for a young black hole, these 
operators can be chosen to act only on the hard and soft modes, 
while for an old black hole, they must also involve early radiation.

In Section~\ref{sec:v2}, which was added after the first version of 
this paper was posted, in response to an intriguing paper by Penington, 
Shenker, Stanford, and Yang~\cite{Penington:2019kki}, we perform 
a similar analysis for a large AdS black hole.  In this case, we 
obtain a result resembling that of Ref.~\cite{Penington:2019kki}:\ 
for an old black hole, we can construct certain operators analogous 
to the interior ones which act only on early radiation.  These 
operators, however, are not relevant for constructing the effective 
interior theory erected at that time; as in the case of a flat 
space black hole, operators relevant for such an effective theory 
must involve both soft modes and early radiation.  A key to 
reconcile these two results and figuring out the relation between 
our construction and entanglement wedge reconstruction of the 
interior~\cite{Penington:2019npb,Almheiri:2019psf,Almheiri:2019hni} 
is time evolution, which we will discuss.

In Section~\ref{sec:concl}, we conclude with a discussion on the relation 
between our findings and the emergence of semiclassical spacetime. 
We conjecture that the existence of (approximately) global operators 
is required for the emergence of the semiclassical picture.

Throughout the paper, we focus on black holes in 4-dimensional spacetime 
that are not significantly rotating or charged, although the restriction 
on specific spacetime dimensions or on non-rotating, non-charged black 
holes is not essential.  We adopt natural units $c = \hbar = 1$, and 
$l_{\rm P}$ denotes the Planck length.

\section{Flat Space Black Hole in a Distant Description}
\label{sec:framework}

In this section and the next, we discuss a flat space (or a small AdS) 
black hole using the framework of Refs.~\cite{Nomura:2018kia,Nomura:2019qps}. 
A key feature of the framework is that the thermal nature of a black hole 
in a distant description can be viewed as arising from entanglement between 
hard and soft modes of low energy quantum fields.%
\footnote{Here and below, low energy fields mean quantum fields existing 
 below the string scale, $1/l_{\rm s}$.}
Modes of a low energy quantum field in the zone region (also called the 
thermal atmosphere)
\begin{equation}
  r_{\rm s} \leq r \leq r_{\rm z}
\label{eq:zone}
\end{equation}
are decomposed into hard and soft modes; the hard modes have frequencies 
$\omega$ and gaps among them $\varDelta\omega$ larger than 
\begin{equation}
  \Delta \approx O\biggl(\frac{1}{M l_{\rm P}^2}\biggr)
\label{eq:Delta}
\end{equation}
as measured in the asymptotic region, while the soft modes have 
$\omega \lesssim \Delta$.  (The hard modes are those which the 
bulk theory can discriminate individually at the quantum level.) 
Here, $r_{\rm z} \approx 3 M l_{\rm P}^2$, and $r_{\rm s}$ is the 
location of the stretched horizon, given by
\begin{equation}
  r_{\rm s} - 2 M l_{\rm P}^2 \sim \frac{l_{\rm s}^2}{M l_{\rm P}^2}.
\label{eq:stretched}
\end{equation}
In a distant description, the classical spacetime picture is applicable 
only outside the stretched horizon, and its location is determined by 
the condition that the proper distance from the mathematical horizon, 
$r = 2 M l_{\rm P}^2$, is of order the string length $l_{\rm s}$.

While the frequencies of the soft modes are small as measured in the 
asymptotic region, their intrinsic dynamical scale is larger at a location 
deeper in the zone, due to large gravitational blueshift.  In particular, 
it is of order the string scale near the stretched horizon, where a majority 
of the modes reside.  (The distribution of the soft modes is given by 
the entropy density that goes as the cubic power of the blueshift factor 
$1/\sqrt{1 - 2Ml_{\rm P}^2/r}$.)  The dynamics of the soft modes there, 
therefore, cannot be described by the low energy theory.%
\footnote{Given that the dynamics is not dictated by the low energy theory, 
 we may call these modes the stretched horizon degrees of freedom instead 
 of soft modes of low energy fields near the stretched horizon.  In fact, 
 the internal dynamics of these modes are expected to be nonlocal in the 
 spatial directions along the horizon~\cite{Hayden:2007cs,Sekino:2008he}.}
It is this dynamics that is responsible for unitarity of the Hawking 
emission process.

The quantity $\Delta$ in Eq.~(\ref{eq:Delta}) is naturally taken to be 
somewhat, e.g.\ by a factor of $O(10)$, larger than the Hawking temperature
\begin{equation}
  T_{\rm H} = \frac{1}{8\pi M l_{\rm P}^2}.
\label{eq:T_H}
\end{equation}
Since $\Delta$ is the inverse timescale for single Hawking emission, the 
uncertainty principle prevents us from specifying the energy of the black 
hole better than that.  Below, we will assume that the energy (mass) of 
a black hole is determined with this maximal precision.  A superposition 
of black holes of masses differing more than $\Delta$ can be treated 
in a straightforward manner.

At a given time $t$, the state of the entire system---with the black hole 
being put in the semiclassical vacuum state---is given by
\begin{equation}
  \ket{\Psi(M)} = \sum_n \sum_{i_n = 1}^{e^{S_{\rm bh}(M-E_n)}} 
    \sum_{a = 1}^{e^{S_{\rm rad}}} c_{n i_n a} \ket{\{ n_\alpha \}} 
    \ket{\psi^{(n)}_{i_n}} \ket{\phi_a}.
\label{eq:sys-state}
\end{equation}
Excitations on a black hole background will be discussed later.  In this 
expression, $\ket{\{ n_\alpha \}}$ are orthonormal states of the hard 
modes, with $n \equiv \{ n_\alpha \}$ representing the set of all occupation 
numbers $n_\alpha$ ($\geq 0$).  The index $\alpha$ collectively denotes 
the species, frequency, and angular-momentum quantum numbers of a mode, 
and $E_n$ is the energy of the state $\ket{\{ n_\alpha \}}$ as measured 
in the asymptotic region (with precision $\Delta$).

$\ket{\psi^{(n)}_{i_n}}$ are orthonormal states of the soft modes entangled 
with $\ket{\{ n_\alpha \}}$ (and hence having energy $M-E_n$ with precision 
$\Delta$).  The density of states for the soft modes is given by the 
Bekenstein-Hawking formula
\begin{equation}
  S_{\rm bh}(M) = 4\pi M^2 l_{\rm P}^2.
\label{eq:S_bh}
\end{equation}
Here, we have assumed that the density of hard mode states is negligible 
compared with that of the soft modes.  This implies that $i_n$ runs over
\begin{equation}
  i_n = 1,\cdots,e^{S_{\rm bh}(M-E_n)}.
\label{eq:i_n}
\end{equation}
Note that with this assumption, the total entropy of the black hole is
\begin{equation}
  \ln\biggl[ \sum_n e^{S_{\rm bh}(M-E_n)} \biggr] 
  \approx \frac{{\cal A}(M)}{4 l_{\rm P}^2},
\label{eq:S-total}
\end{equation}
where ${\cal A}(M) = 16\pi M^2 l_{\rm P}^4$ is the area of the black hole, 
reproducing the standard interpretation of the Bekenstein-Hawking entropy.
The last factor $\ket{\phi_a}$ in Eq.~(\ref{eq:sys-state}) represents the set 
of orthonormal states representing the system in the far region $r > r_{\rm z}$. 

By the black hole vacuum, we mean that there is no physical excitation 
identifiable at the semiclassical level.  This implies that any attribute 
a hard mode state may have is compensated by that of the corresponding 
soft mode states (within the precision allowed by the uncertainty principle). 
In particular, this implies that soft mode states associated with different 
hard mode states are orthogonal:
\begin{equation}
  \inner{\psi^{(m)}_{i_m}}{\psi^{(n)}_{j_n}} = \delta_{m n} \delta_{i_m j_n}.
\label{eq:soft-orthonorm}
\end{equation}
We also take the states in the far region, $\ket{\phi_a}$, to be given by 
those of Hawking radiation emitted earlier, i.e.\ emitted from $r \approx 
r_{\rm z}$ to the asymptotic region before time $t$.  $S_{\rm rad}$ 
in Eq.~(\ref{eq:sys-state}) is then the coarse-grained entropy of this 
early radiation.

We take the state in Eq.~(\ref{eq:sys-state}) to be normalized:
\begin{equation}
  \sum_n \sum_{i_n = 1}^{e^{S_{\rm bh}(M-E_n)}} \sum_{a = 1}^{e^{S_{\rm rad}}} 
    |c_{n i_n a}|^2 = 1.
\label{eq:norm}
\end{equation}
We also assume that the ultraviolet dynamics near the stretched horizon is 
chaotic, well scrambling the black hole state~\cite{Hayden:2007cs,Sekino:2008he}. 
In particular, we assume that the coefficients $c_{n i_n a}$ take generic 
values in the spaces of the hard and soft modes.  This implies that 
statistically
\begin{equation}
  |c_{n i_n a}| \sim \frac{1}{\sqrt{S_{\rm tot}}},
\label{eq:c-gen}
\end{equation}
where
\begin{equation}
  S_{\rm tot} 
  \equiv \biggl( \sum_n e^{S_{\rm bh}(M-E_n)} \biggr)\, e^{S_{\rm rad}} 
  = \biggl( \sum_n e^{-\frac{E_n}{T_{\rm H}}} \biggr)\, 
    e^{S_{\rm bh}(M)} e^{S_{\rm rad}}.
\label{eq:S_tot}
\end{equation}
The standard thermal nature of the black hole is then obtained upon tracing 
out the soft modes:
\begin{equation}
  {\rm Tr}_{\rm soft} \ket{\Psi(M)} \bra{\Psi(M)}
  = \frac{1}{\sum_m e^{-\frac{E_m}{T_{\rm H}}}} \sum_n 
    e^{-\frac{E_n}{T_{\rm H}}} \ket{\{ n_\alpha \}} \bra{\{ n_\alpha \}} 
    \otimes \rho_{\phi,n},
\label{eq:rho_HR}
\end{equation}
where $\rho_{\phi,n}$ are reduced density matrices for the early radiation, 
whose $n$-dependence is small and of order $1/\sqrt{e^{S_{\rm bh}(M)}}$. 
Note that in order to obtain the correct Boltzmann factor, $\propto 
e^{-E_n/T_{\rm H}}$, it is essential that the coefficients $c_{n i_n a}$ 
take generic values across all low energy species, i.e.\ $n$ runs over 
all low energy species~\cite{Nomura:2019qps}.

In a distant description, the system of a black hole and radiation evolves 
unitarily with the state taking the form of Eq.~(\ref{eq:sys-state}) at each 
moment in time.  In particular, the entanglement entropy between the black 
hole and radiation
\begin{equation}
  S^{\rm vN}_{\rm hard + soft} = S^{\rm vN}_{\rm rad}
\end{equation}
follows the Page curve~\cite{Page:1993wv}, where $S^{\rm vN}_A$ is the von~Neumann 
entropy of subsystem $A$.  Throughout the history of the black hole, the number 
of hard modes is much smaller than that of the soft modes.  (Note that we 
are only interested in states that do not yield significant backreaction on 
spacetime, which limits the number of possible hard mode states.)  Furthermore, 
the coarse-grained entropies of the soft modes and radiation are both of 
order $M^2 l_{\rm P}^2$, except for the very beginning and end of the black 
hole evolution.  We therefore have
\begin{equation}
  \ln {\rm dim}{\cal H}_{\rm hard} \ll 
    S_{\rm bh}(M-E_n), S_{\rm rad} \approx O(M^2 l_{\rm P}^2).
\label{eq:rel-sizes}
\end{equation}
We stress that this relation holds both before and after the Page time, at 
which the coarse-grained entropy of the radiation becomes approximately equal 
to that of the black hole.

Incidentally, by performing the Schmidt decomposition in the space of soft-mode 
and radiation states for each $n$, the state in Eq.~(\ref{eq:sys-state}) can be 
written as
\begin{equation}
  \ket{\Psi(M)} = \sum_n \sum_{i_n=1}^{{\cal N}_n} 
    c^n_{i_n} \ket{H_n} \ket{S_{n,i_n}} \ket{R_{n,i_n}},
\label{eq:ent-str}
\end{equation}
where $\ket{H_n}$, $\ket{S_{n,i_n}}$, and $\ket{R_{n,i_n}}$ are states of 
the hard modes, soft modes, and radiation, respectively, and
\begin{equation}
  {\cal N}_n = {\rm min}\{ e^{S_{\rm bh}(M-E_n)}, e^{S_{\rm rad}} \}.
\label{eq:N_n}
\end{equation}
This expression elucidates why the entanglement argument for 
firewalls~\cite{Almheiri:2012rt} does not apply here.  The entanglement 
responsible for unitarity has to do with the summations of indices 
$i_n$ (in fact, predominantly the vacuum index $i_0$) shared between 
the soft-mode and radiation states, while the entanglement necessary 
for the interior spacetime (see below) has to do with the index $n$. 
These two are compatible because the number of terms associated with 
the sum over $n$ is much smaller than ${\cal N}_n$.

Let us now discuss excitations.  A small excitation composed of 
constituents with $\omega \gtrsim \Delta$  in the zone can be described 
by annihilation and creation operators acting on the hard modes
\begin{align}
  b_\gamma &= \sum_n \sqrt{n_\gamma}\, 
    \ket{\{ n_\alpha - \delta_{\alpha\gamma} \}} \bra{\{ n_\alpha \}},
\label{eq:ann}\\
  b_\gamma^\dagger &= \sum_n \sqrt{n_\gamma + 1}\, 
    \ket{\{ n_\alpha + \delta_{\alpha\gamma} \}} \bra{\{ n_\alpha \}}.
\label{eq:cre}
\end{align}
In particular, a small object falling from the far region into the black 
hole, which has the characteristic size $d$ in the angular directions 
much smaller than the horizon, i.e.\ $d \ll M l_{\rm P}^2$, can be 
described in this manner.

In a distant description, a small object falling into the black hole is 
absorbed into the stretched horizon when it reaches there, whose information 
will be later sent back to ambient space by Hawking emission.  This description, 
however, is not useful for addressing the question of what the falling object 
will actually see.  Because of the large discrepancy between the stationary 
frame and the frame of the object, macroscopic time experienced by the 
object is mapped to an extremely short time for a stationary observer at 
the location of the object.  In particular, anything the object experiences 
inside the horizon occurs almost ``instantaneously'' for a stationary observer 
at $r = r_{\rm s}$.  Understanding an object's experiences, therefore, requires 
time evolution different from the distant one, specifically an evolution 
associated with the proper time of the object.

\section{Effective Theory of the Black Hole Interior}
\label{sec:interior}

The effective theory describing the black hole interior can be erected {\it at 
each time $t$} by coarse-graining the soft modes and radiation:\ the degrees 
of freedom that cannot be resolved by a fallen object in the timescale available 
to it.  Suppose that the state of the system at time $t$ (with the black 
hole put in the semiclassical vacuum) is given by Eq.~(\ref{eq:sys-state}) 
in a distant description.  We can then define a set of coarse-grained states 
each of which is entangled with a specific hard mode state:
\begin{equation}
  \ketc{\{ n_\alpha \}} 
  \propto \sum_{i_n = 1}^{e^{S_{\rm bh}(M-E_n)}} \sum_{a = 1}^{e^{S_{\rm rad}}} 
    c_{n i_n a} \ket{\psi^{(n)}_{i_n}} \ket{\phi_a},
\label{eq:coarse-prop}
\end{equation}
where we have used the same label as the corresponding hard mode state to 
specify the coarse-grained state, which we denote by the double ket symbol. 
Note that the meaning of the coarse-graining here is that the state 
$\ketc{\{ n_\alpha \}}$ on the left-hand side corresponds to multiple 
different microstates on the right-hand side, depending on the state of 
the black hole and radiation represented by the coefficients $c_{n i_n a}$.

Using Eq.~(\ref{eq:c-gen}), we find that the squared norm of the (non-normalized) 
state on the right-hand side of Eq.~(\ref{eq:coarse-prop}) is given by
\begin{equation}
  \sum_{i_n = 1}^{e^{S_{\rm bh}(M-E_n)}} 
    \sum_{a = 1}^{e^{S_{\rm rad}}} |c_{n i_n a}|^2 
 = \frac{e^{-\frac{E_n}{T_{\rm H}}}}
    {\Bigl( \sum_m e^{-\frac{E_m}{T_{\rm H}}} \Bigr)} 
    \left[ 1 + O\biggl( \frac{1}{\sqrt{e^{S_{\rm bh}(M-E_n)} e^{S_{\rm rad}}}} 
      \biggr) \right]
\label{eq:coarse-norm}
\end{equation}
for generic black hole and radiation microstates.  Here, the second term 
in the square brackets represents the size of statistical fluctuations 
over different microstates.  Therefore, the normalized coarse-grained state 
$\ketc{\{ n_\alpha \}}$ is given for generic microstates by
\begin{equation}
  \ketc{\{ n_\alpha \}} 
  = e^{\frac{E_n}{2 T_{\rm H}}} \sqrt{\sum_m e^{-\frac{E_m}{T_{\rm H}}}}\, 
    \sum_{i_n = 1}^{e^{S_{\rm bh}(M-E_n)}} \sum_{a = 1}^{e^{S_{\rm rad}}} 
    c_{n i_n a} \ket{\psi^{(n)}_{i_n}} \ket{\phi_a},
\label{eq:coarse}
\end{equation}
up to a fractional correction of order $1/e^{\# M^2 l_{\rm P}^2}$ in the 
overall normalization, where $\#$ is a number that does not depend on 
$M l_{\rm P}$.

The state in the effective theory corresponding to the state in 
Eq.~(\ref{eq:sys-state}) can be written in terms of the coarse-grained 
states in Eq.~(\ref{eq:coarse}) as
\begin{equation}
  \ketc{\Psi(M)} 
  = \frac{1}{\sqrt{\sum_m e^{-\frac{E_m}{T_{\rm H}}}}} 
    \sum_n e^{-\frac{E_n}{2 T_{\rm H}}} 
    \ket{\{ n_\alpha \}} \ketc{\{ n_\alpha \}},
\label{eq:BH-coarse}
\end{equation}
regardless of the values of $c_{n i_n a}$.  This takes the form of 
the standard thermofield double state in the two-sided black hole 
picture~\cite{Unruh:1976db,Israel:1976ur}, although $\ket{\{ n_\alpha \}}$ 
here represent the states only of the hard modes.  We emphasize 
that in order to obtain the correct Boltzmann-weight coefficients, 
$\propto e^{-E_n/2 T_{\rm H}}$, it is important that the 
black hole has soft modes with the density of states given 
by $e^{S_{\rm bh}(E_{\rm soft})}$, and that the hard and soft 
modes are well scrambled, giving $c_{n i_n a}$ that take values 
statistically independent of $n$.  This coarse-graining leads 
to the apparent uniqueness of the infalling vacuum, despite the 
existence of exponentially many black hole microstates.

One can now define the annihilation and creation operators $\tilde{b}_\gamma$ 
and $\tilde{b}_\gamma^\dagger$ acting on the coarse-grained states as
\begin{align}
  \tilde{b}_\gamma &= \sum_n \sqrt{n_\gamma}\, 
    \ketc{\{ n_\alpha - \delta_{\alpha\gamma} \}} 
    \brac{\{ n_\alpha \}},
\label{eq:ann-m}\\
  \tilde{b}_\gamma^\dagger &= \sum_n \sqrt{n_\gamma + 1}\, 
    \ketc{\{ n_\alpha + \delta_{\alpha\gamma} \}} 
    \brac{\{ n_\alpha \}}
\label{eq:cre-m}
\end{align}
{\it at the level of the effective theory}.  The annihilation and creation 
operators relevant for an infalling observer can then be given by
\begin{align}
  a_\xi &= \sum_\gamma 
    \bigl( \alpha_{\xi\gamma} b_\gamma 
    + \beta_{\xi\gamma} b_\gamma^\dagger 
    + \zeta_{\xi\gamma} \tilde{b}_\gamma 
    + \eta_{\xi\gamma} \tilde{b}_\gamma^\dagger \bigr),
\label{eq:a_xi}\\
  a_\xi^\dagger &= \sum_\gamma 
    \bigl( \beta_{\xi\gamma}^* b_\gamma 
    + \alpha_{\xi\gamma}^* b_\gamma^\dagger 
    + \eta_{\xi\gamma}^* \tilde{b}_\gamma 
    + \zeta_{\xi\gamma}^* \tilde{b}_\gamma^\dagger \bigr),
\label{eq:a_xi-dag}
\end{align}
where $b_\gamma$ and $b_\gamma^\dagger$ are the operators in 
Eqs.~(\ref{eq:ann},~\ref{eq:cre}), $\xi$ is the label in which 
the frequency $\omega$ with respect to $t$ is traded with the frequency 
$\Omega$ associated with the infalling time, and $\alpha_{\xi\gamma}$, 
$\beta_{\xi\gamma}$, $\zeta_{\xi\gamma}$, and $\eta_{\xi\gamma}$ are the 
Bogoliubov coefficients calculable using the standard field theory method. 
The generator of time evolution in the infalling description is then 
given by
\begin{equation}
  H = \sum_\xi \Omega a_\xi^\dagger a_\xi 
    + H_{\rm int}\bigl( a_\xi, a_\xi^\dagger \bigr).
\label{eq:H_eff}
\end{equation}
This leads to the physics of a smooth horizon.  The existence of the 
operators $a_\xi$ and $a_\xi^\dagger$ implies that there is a subsector 
in the original microscopic theory encoding the experience of an object 
after it crosses the horizon.

How can the operators $\tilde{b}_\gamma$ and $\tilde{b}_\gamma^\dagger$ 
be constructed {\it at the microscopic level}?  One way is simply to 
use Eq.~(\ref{eq:coarse}) in the expression in Eqs.~(\ref{eq:ann-m},%
~\ref{eq:cre-m}):
\begin{align}
  \tilde{b}_\gamma &= \left( \sum_m e^{-\frac{E_m}{T_{\rm H}}} \right) 
    \sum_n \sqrt{n_\gamma}\, e^{\frac{E_{n_-}+E_n}{2T_{\rm H}}}
\nonumber\\
  & \quad\times
    \sum_{i_{n_-} = 1}^{e^{S_{\rm bh}(M-E_{n_-})}} 
    \sum_{j_n = 1}^{e^{S_{\rm bh}(M-E_n)}} 
    \sum_{a = 1}^{e^{S_{\rm rad}}} \sum_{b = 1}^{e^{S_{\rm rad}}} 
    c_{n_- i_{n_-} a} c^*_{n j_n b}\, 
    \ket{\psi^{(n_-)}_{i_{n_-}}} \ket{\phi_a} \bra{\psi^{(n)}_{j_n}} \bra{\phi_b},
\label{eq:ann-m-orig}\\
  \tilde{b}_\gamma^\dagger &= \left( \sum_m e^{-\frac{E_m}{T_{\rm H}}} \right) 
    \sum_n \sqrt{n_\gamma + 1}\, e^{\frac{E_{n_+}+E_n}{2T_{\rm H}}}
\nonumber\\
  & \quad\times
    \sum_{i_{n_+} = 1}^{e^{S_{\rm bh}(M-E_{n_+})}} 
    \sum_{j_n = 1}^{e^{S_{\rm bh}(M-E_n)}} 
    \sum_{a = 1}^{e^{S_{\rm rad}}} \sum_{b = 1}^{e^{S_{\rm rad}}} 
    c_{n_+ i_{n_+} a} c^*_{n j_n b}\, 
    \ket{\psi^{(n_+)}_{i_{n_+}}} \ket{\phi_a} \bra{\psi^{(n)}_{j_n}} \bra{\phi_b},
\label{eq:cre-m-orig}
\end{align}
where $n_\pm \equiv \{ n_\alpha \pm \delta_{\alpha\gamma} \}$, 
and $E_{n_\pm}$ are the energies of the hard mode states 
$\ket{\{ n_\alpha \pm \delta_{\alpha\gamma} \}}$ as measured in 
the asymptotic region.  These operators can play the role of annihilation 
and creation operators in the space spanned by the coarse-grained states. 
In particular, their matrix elements are
\begin{align}
  \brac{\{ \kappa_\alpha \}} \tilde{b}_\gamma \ketc{\{ \lambda_\alpha \}} 
  &= \sqrt{\lambda_\gamma}\,\, 
    \delta_{\{ \kappa_\alpha \} \{ \lambda_\alpha - \delta_{\alpha\gamma} \}},
\label{eq:ann-mat}\\
  \brac{\{ \kappa_\alpha \}} \tilde{b}_\gamma^\dagger \ketc{\{ \lambda_\alpha \}} 
  &= \sqrt{\lambda_\gamma + 1}\,\, 
    \delta_{\{ \kappa_\alpha \} \{ \lambda_\alpha + \delta_{\alpha\gamma} \}},
\label{eq:cre-mat}
\end{align}
up to corrections of order $1/e^{\# M^2 l_{\rm P}^2}$.  It is important to 
notice, however, that these operators do not satisfy the exact algebra of 
annihilation and creation operators at the microscopic level.  Indeed,
\begin{equation}
  [\tilde{b}_\beta, \tilde{b}_\gamma^\dagger] = \delta_{\beta \gamma} 
    \left( \sum_m e^{-\frac{E_m}{T_{\rm H}}} \right) 
    \sum_n e^{\frac{E_n}{T_{\rm H}}} 
    \sum_{i_n = 1}^{e^{S_{\rm bh}(M-E_n)}} \sum_{j_n = 1}^{e^{S_{\rm bh}(M-E_n)}} 
    \sum_{a = 1}^{e^{S_{\rm rad}}} \sum_{b = 1}^{e^{S_{\rm rad}}} 
    \ket{\psi^{(n)}_{i_n}} \ket{\phi_a} \bra{\psi^{(n)}_{j_n}} \bra{\phi_b},
\label{eq:algebra}
\end{equation}
which is not the identity operator for $\beta = \gamma$.  It is only 
in the space of coarse-grained states that these operators obey the algebra 
of annihilation and creation operators:
\begin{equation}
  \brac{\{ \kappa_\alpha \}} [\tilde{b}_\beta, \tilde{b}_\gamma^\dagger] 
    \ketc{\{ \lambda_\alpha \}} = \delta_{\beta \gamma},
\label{eq:algebra-mat-1}
\end{equation}
\begin{equation}
  \brac{\{ \kappa_\alpha \}} [\tilde{b}_\beta, \tilde{b}_\gamma] 
    \ketc{\{ \lambda_\alpha \}} 
  =  \brac{\{ \kappa_\alpha \}} [\tilde{b}_\beta^\dagger, \tilde{b}_\gamma^\dagger] 
    \ketc{\{ \lambda_\alpha \}} 
  = 0,
\label{eq:algebra-mat-2}
\end{equation}
which have corrections only of order $1/e^{\# M^2 l_{\rm P}^2}$.

Can other microscopic operators be chosen as the annihilation and 
creation operators in the effective theory?  One might think that 
any operators mapping a generic microstate of $\ketc{\{ n_\alpha \}}$ 
(i.e.\ a state in Eq.~(\ref{eq:coarse}) with generic $c_{n i_n a}$) 
to those of $\ketc{\{ n_\alpha - \delta_{\alpha\gamma} \}}$ 
and $\ketc{\{ n_\alpha + \delta_{\alpha\gamma} \}}$ would work 
as $\tilde{b}_\gamma$ and $\tilde{b}_\gamma^\dagger$, respectively. 
This is, however, not the case.  Since a single coarse-grained state 
$\ketc{\{ n_\alpha \}}$ corresponds to many microstates, the state 
obtained by acting such generic operators to a specific microstate, 
in particular the state $\ketc{\{ n_\alpha \}}$ involving the 
specific coefficients $c_{n i_n a}$ appearing in the state of 
the system in Eq.~(\ref{eq:sys-state}), may not have appropriate 
inner products with the corresponding states $\ketc{\{ m_\alpha \}}$'s 
($\{ m_\alpha \} \neq \{ n_\alpha \}$) involving the same $c_{n i_n a}$. 
This would mean that those microscopic operators do not serve as 
annihilation and creation operators in the effective theory erected 
on the state having these specific coefficients $c_{n i_n a}$.

As an example, consider the set of candidate operators
\begin{equation}
  \tilde{b}_\gamma \stackrel[]{?}{=} 
    c \sum_n \sqrt{n_\gamma}\, \ket{f^{(n_-)}} \bra{g^{(n)}},
\qquad
  \tilde{b}_\gamma^\dagger \stackrel[]{?}{=} 
    c^* \sum_n \sqrt{n_\gamma + 1}\, \ket{g^{(n_+)}} \bra{f^{(n)}},
\label{eq:candidates}
\end{equation}
where $c$ is a normalization constant, and
\begin{equation}
  \ket{f^{(n)}} = \sum_{i_n = 1}^{e^{S_{\rm bh}(M-E_n)}} 
    f_{n i_n} \ket{\psi^{(n)}_{i_n}},
\qquad
  \ket{g^{(n)}} = \sum_{i_n = 1}^{e^{S_{\rm bh}(M-E_n)}} 
    g_{n i_n} \ket{\psi^{(n)}_{i_n}}
\label{eq:generic-ms}
\end{equation}
with generic coefficients satisfying $\sum_{i_n} |f_{n i_n}|^2 = 
\sum_{i_n} |g_{n i_n}|^2 = 1$.  This gives
\begin{equation}
  \brac{\{ \kappa_\alpha \}} \tilde{b}_\gamma 
    \ketc{\{ \lambda_\alpha \}} = c \sqrt{\lambda_\gamma}\, 
    \delta_{\{ \kappa_\alpha \} \{ \lambda_\alpha - \delta_{\alpha\gamma} \}} 
    O\biggl(\frac{1}{\sqrt{e^{S_{\rm bh}(M-E_\kappa)} e^{S_{\rm bh}(M-E_\lambda)} 
      e^{S_{\rm rad}}}}\biggr),
\label{eq:contrad-1}
\end{equation}
\begin{equation}
  \brac{\{ \kappa_\alpha \}} \tilde{b}_\gamma \tilde{b}_\gamma^\dagger 
    \ketc{\{ \lambda_\alpha \}} 
  = |c|^2 (\lambda_\gamma + 1)\, \delta_{\{ \kappa_\alpha \} \{ \lambda_\alpha \}} 
        O\biggl(\frac{1}{e^{S_{\rm bh}(M-E_\lambda)}}\biggr),
\label{eq:contrad-2}
\end{equation}
and there is no choice of $c$ that can make both of these relations 
compatible with the algebra in the effective theory.

The consideration above provides an argument for the necessity 
of the dependence of the microscopic operators $\tilde{b}_\gamma$ 
and $\tilde{b}_\gamma^\dagger$ on the state of the system, in 
particular $c_{n i_n a}$ in Eq.~(\ref{eq:sys-state}).  This, 
however, still allows for operators other than those in 
Eqs.~(\ref{eq:ann-m-orig},~\ref{eq:cre-m-orig}).

Let us consider the operators
\begin{align}
  \tilde{b}_\gamma &= c \sum_n \sqrt{n_\gamma}\, 
    e^{\frac{E_{n_-} + E_n}{2 T_{\rm H}}} 
    \sum_{i_{n_-} = 1}^{e^{S_{\rm bh}(M-E_{n_-})}} 
    \sum_{j_n = 1}^{e^{S_{\rm bh}(M-E_n)}} 
    \sum_{a = 1}^{e^{S_{\rm rad}}} c_{n_- i_{n_-} a} c_{n j_n a}^*\, 
    \ket{\psi^{(n_-)}_{i_{n_-}}} \bra{\psi^{(n)}_{j_n}},
\label{eq:ann-av}\\
  \tilde{b}_\gamma^\dagger &= c \sum_n \sqrt{n_\gamma + 1}\, 
    e^{\frac{E_{n_+} + E_n}{2 T_{\rm H}}} 
    \sum_{i_{n_+} = 1}^{e^{S_{\rm bh}(M-E_{n_+})}} 
    \sum_{j_n = 1}^{e^{S_{\rm bh}(M-E_n)}} 
    \sum_{a = 1}^{e^{S_{\rm rad}}} c_{n_+ i_{n_+} a} c_{n j_n a}^*\, 
    \ket{\psi^{(n_+)}_{i_{n_+}}} \bra{\psi^{(n)}_{j_n}},
\label{eq:cre-av}
\end{align}
where $c$ is a real number.  Note that the combinations of $c_{n j_n a}$'s 
appearing here, $\sum_{a = 1}^{e^{S_{\rm rad}}} c_{m i_m a} c_{n j_n a}^*$, 
are those in the reduced density matrix for the hard and soft modes
\begin{equation}
  {\rm Tr}_{\rm rad} \ket{\Psi(M)} \bra{\Psi(M)}
  = \sum_m \sum_n \sum_{i_m = 1}^{e^{S_{\rm bh}(M-E_m)}} 
    \sum_{j_n = 1}^{e^{S_{\rm bh}(M-E_n)}} 
    \sum_{a = 1}^{e^{S_{\rm rad}}} c_{m i_m a} c_{n j_n a}^*\, 
    \ket{\{ m_\alpha \}} \ket{\psi^{(m)}_{i_m}} 
    \bra{\{ n_\alpha \}} \bra{\psi^{(n)}_{j_n}},
\label{eq:rho_HS}
\end{equation}
so that they can be determined purely from the state in the black hole region.

With this choice of $\tilde{b}_\gamma$ and $\tilde{b}_\gamma^\dagger$, we obtain
\begin{align}
  \brac{\{ \kappa_\alpha \}} \tilde{b}_\gamma 
    \ketc{\{ \lambda_\alpha \}} 
  &= \frac{c}{e^{S_{\rm rad}} \sum_m e^{-\frac{E_m}{T_{\rm H}}}} 
    \sqrt{\lambda_\gamma}\, 
    \delta_{\{ \kappa_\alpha \} \{ \lambda_\alpha - \delta_{\alpha\gamma} \}},
\label{eq:eff-1}\\
  \brac{\{ \kappa_\alpha \}} \tilde{b}_\gamma^\dagger 
    \ketc{\{ \lambda_\alpha \}} 
  &= \frac{c}{e^{S_{\rm rad}} \sum_m e^{-\frac{E_m}{T_{\rm H}}}} 
    \sqrt{\lambda_\gamma + 1}\, 
    \delta_{\{ \kappa_\alpha \} \{ \lambda_\alpha + \delta_{\alpha\gamma} \}},
\label{eq:eff-2}
\end{align}
and
\begin{align}
  \brac{\{ \kappa_\alpha \}} \tilde{b}_\gamma \tilde{b}_\gamma^\dagger 
    \ketc{\{ \lambda_\alpha \}} 
  &= \left( \frac{c}{e^{S_{\rm rad}} \sum_m e^{-\frac{E_m}{T_{\rm H}}}} \right)^2 
    (\lambda_\gamma + 1)\, \delta_{\{ \kappa_\alpha \} \{ \lambda_\alpha \}} \, 
    \left[ 1 + O\Biggl(\frac{e^{S_{\rm rad}}}
      {e^{S_{\rm bh}(M-E_\lambda)}}\Biggr) \right],
\label{eq:eff-3}\\
  \brac{\{ \kappa_\alpha \}} \tilde{b}_\gamma^\dagger \tilde{b}_\gamma 
    \ketc{\{ \lambda_\alpha \}} 
  &= \left( \frac{c}{e^{S_{\rm rad}} \sum_m e^{-\frac{E_m}{T_{\rm H}}}} \right)^2 
    \lambda_\gamma\, \delta_{\{ \kappa_\alpha \} \{ \lambda_\alpha \}} \, 
    \left[ 1 + O\Biggl(\frac{e^{S_{\rm rad}}}
      {e^{S_{\rm bh}(M-E_\lambda)}}\Biggr) \right],
\label{eq:eff-4}
\end{align}
up to corrections of order $1/e^{\# M^2 l_{\rm P}^2}$.  We thus find that for
\begin{equation}
  e^{S_{\rm rad}} \ll e^{S_{\rm bh}(M-E_\lambda)} \approx e^{S_{\rm bh}(M)},
\label{eq:before-Page}
\end{equation}
the second terms in the square brackets in Eqs.~(\ref{eq:eff-3},~\ref{eq:eff-4}) 
are negligible, so that the $\tilde{b}_\gamma$ and $\tilde{b}_\gamma^\dagger$ 
with
\begin{equation}
  c = e^{S_{\rm rad}} \sum_m e^{-\frac{E_m}{T_{\rm H}}}
\label{eq:c}
\end{equation}
can play the role of the annihilation and creation operators in the effective 
theory.  In fact, we can show that these operators also satisfy the required 
commutation relations in Eqs.~(\ref{eq:algebra-mat-1},~\ref{eq:algebra-mat-2}) 
when the condition in Eq.~(\ref{eq:before-Page}) is satisfied, namely when 
the black hole is young.

On the other hand, if the black hole is old, i.e.\ $e^{S_{\rm rad}} \gg 
e^{S_{\rm bh}(M)}$, then the second terms in the square brackets dominate in 
Eqs.~(\ref{eq:eff-3},~\ref{eq:eff-4}), jeopardizing the possibility for the 
$\tilde{b}_\gamma$ and $\tilde{b}_\gamma^\dagger$ to serve as the annihilation 
and creation operators in the effective theory for any choice of $c$. 
Technically, this is because the factor obtained by acting $b_\gamma^\dagger$ 
or $b_\gamma$ to $\ketc{\{ \lambda_\alpha \}}$
\begin{equation}
  \sum_{i_{\lambda_\pm} = 1}^{e^{S_{\rm bh}(M-E_{\lambda_\pm})}} 
  \sum_{j_\lambda = 1}^{e^{S_{\rm bh}(M-E_\lambda)}} 
  \sum_{a = 1}^{e^{S_{\rm rad}}} \sum_{b = 1}^{e^{S_{\rm rad}}} 
    c_{\lambda_\pm i_{\lambda_\pm} a} c_{\lambda j_\lambda a}^* 
    c_{\lambda j_\lambda b} 
  \ket{\psi^{(\lambda_\pm)}_{i_{\lambda_\pm}}} \ket{\phi_b}
\end{equation}
is dominated by the $a=b$ terms if and only if the condition in 
Eq.~(\ref{eq:before-Page}) is met, giving the state proportional to
\begin{equation}
  \sum_{i_{\lambda_\pm} = 1}^{e^{S_{\rm bh}(M-E_{\lambda_\pm})}} 
  \sum_{a = 1}^{e^{S_{\rm rad}}} c_{\lambda_\pm i_{\lambda_\pm} a} 
  \ket{\psi^{(\lambda_\pm)}_{i_{\lambda_\pm}}} \ket{\phi_a},
\end{equation}
and hence to the $\ketc{\{ \lambda_\alpha \pm \delta_{\alpha\gamma} \}}$ 
obtained using the specific $c_{n j_n a}$'s appearing in the state 
of the system.  This shows how the Page time can be relevant in 
the construction of the interior operators, despite the fact that 
the hard mode and radiation states take a separable form as in 
Eq.~(\ref{eq:rho_HR}) throughout the history of the black hole.%
\footnote{This separation is not a statistical statement; 
 i.e., it does not receive exponentially small corrections from 
 statistics.  A similar statement applies for a state of the form 
 in Eq.~(\ref{eq:sys-large}) in the next section.}
In fact, for an old black hole, we do not see how one can construct 
the appropriate annihilation and creation operators in the effective 
theory using only the information in the black hole reduced density 
matrix in Eq.~(\ref{eq:rho_HS}).

Incidentally, a construction of $\tilde{b}_\gamma$ and 
$\tilde{b}_\gamma^\dagger$ involving only radiation states 
is not possible.  What allowed the construction of operators in 
Eqs.~(\ref{eq:ann-av},~\ref{eq:cre-av}) is the correlations between 
the attributes of the hard and soft modes coming from the constraints 
imposed on the black hole vacuum state (the requirement that it does 
not have any features associated with semiclassical excitations). 
Such correlations do not exist between the hard modes and radiation.

As discussed in Ref.~\cite{Nomura:2018kia}, the effective theory of the 
interior erected as above describes only a limited spacetime region:\ 
the causal domain of the union of the zone and its mirror region 
on the spatial hypersurface at $t$ (the time at which the effective 
theory is erected) in the effective two-sided geometry.  The black 
hole singularity may be regarded as a manifestation of the fact 
that this theory is obtained by coarse-graining and hence represents 
a finite-dimensional, non-unitary system.  Specific operators used 
in Eqs.~(\ref{eq:a_xi},~\ref{eq:a_xi-dag}), for example those in 
Eqs.~(\ref{eq:ann-m-orig},~\ref{eq:cre-m-orig}), are selected presumably 
because they correspond to observables which classicalize within 
such a finite-dimensional system~\cite{Nomura:2019qps}.  Locality 
seems to play a key role in this quantum-to-classical transition.

The fact that an effective theory represents only a limited spacetime 
region implies that the picture of the whole interior, as described by 
general relativity, can be obtained only by using multiple effective 
theories erected at different times.  This is the sense in which the 
global spacetime in general relativity emerges from the microscopic 
description.

\subsubsection*{Relation to the work by Papadodimas and Raju}

A construction of interior operators similar to the one described here, 
based on the doubled Hilbert space structure, was considered in the 
well-known work by Papadodimas and Raju~\cite{Papadodimas:2012aq,%
Papadodimas:2013jku,Papadodimas:2015jra}.  While the mathematical 
structures of the two are related (at the level of dividing the 
system into two components), their physical implementations are different 
in several key aspects, leading to different solutions to the firewall 
paradoxes~\cite{Almheiri:2012rt,Almheiri:2013hfa,Marolf:2013dba}.

In Ref.~\cite{Papadodimas:2012aq}, the doubled Hilbert space structure 
was obtained by coarse-graining bulk fields (generalized free fields 
in CFT) to separate the degrees of freedom represented by one of the 
Hilbert space factors, which increase as the black hole evaporates. 
(Note that this coarse-graining is different from that discussed in 
Eqs.~(\ref{eq:coarse-prop}~--~\ref{eq:coarse}).)  In contrast, our 
hard modes are selected from the black hole degrees of freedom by 
an energetic criterion, which thus decrease as the evaporation 
progresses.  These modes have gaps larger than the Hawking temperature 
(instead of the effectively continuous spectrum envisioned in 
Ref.~\cite{Papadodimas:2012aq}), which plays an important role 
in the picture as we have seen in this and the previous sections. 
This leads, for example, to an entanglement structure different from 
that envisioned in the earlier work; in particular, the hard mode 
and radiation states take a separable form throughout the history 
of the black hole.

In Refs.~\cite{Papadodimas:2013jku,Papadodimas:2015jra}, which are supposed 
to subsume the earlier construction in Ref.~\cite{Papadodimas:2012aq}, 
the ``state-dependence'' of the map between boundary operators and 
bulk local operators was invoked to address the firewall paradoxes. 
Problems of the state-dependence---the frozen vacuum and Born 
rule problems---were discussed in Ref.~\cite{Bousso:2013ifa} and 
Ref.~\cite{Marolf:2015dia}, respectively.  In our construction, interior 
operators do depend on the microstate on which they are built, in the sense 
that operators $\tilde{b}_\gamma$ and $\tilde{b}_\gamma^\dagger$ depend 
on $c_{n i_n a}$ as shown in Eqs.~(\ref{eq:ann-m-orig},~\ref{eq:cre-m-orig}) 
and Eqs.~(\ref{eq:ann-av},~\ref{eq:cre-av}).  As discussed in 
Ref.\cite{Nomura:2019qps}, however, the fact that our hard modes 
are selected energetically (which is also the case in a large AdS 
black hole as discussed in the next section) allows us to avoid 
the problems in Refs.~\cite{Bousso:2013ifa,Marolf:2015dia}.

Specifically, consider the space, ${\cal H}_M$, of pure states in which 
the energy $E$ in a spatial region is bounded by $E < M$, where $M$ 
is sufficiently large that a typical state in ${\cal H}_M$ is a black 
hole state.  We then consider the space of all states that are obtained 
by acting appropriately smoothed hard mode (semiclassical) operators 
on any of the black hole microstates in ${\cal H}_M$ and have energies 
smaller than $M + \delta E$.  This space, denoted by $B_{\delta E} 
{\cal H}_M$, has dimension $e^{S_{\rm bh}(M) + S_{\rm exc}}$, 
where $S_{\rm exc}$ is the entropy of the possible semiclassical 
excitations.  One can then show that a typical state $\ket{\psi}$ 
in ${\cal H}_{M + \delta E}$ can be written as
\begin{equation}
  \ket{\psi} = \sin\theta\, \ket{\psi_{\rm exc}} 
    + \cos\theta\, \ket{\psi_{\rm vac}}
\label{eq:decomp}
\end{equation}
with
\begin{equation}
  \sin^2\!\theta \sim 
    e^{-\left( \frac{\delta E}{T_{\rm H}} - S_{\rm exc} \right)}.
\label{eq:atypical}
\end{equation}
Here, $\ket{\psi_{\rm exc}}$ and $\ket{\psi_{\rm vac}}$ are elements of 
$B_{\delta E} {\cal H}_M$ and its complement ${\cal H}_{M + \delta E} / 
B_{\delta E} {\cal H}_M$, respectively, and $T_{\rm H} = 1/8\pi M l_{\rm P}^2$. 
Assuming that semiclassical excitations are well within the Bekenstein 
bound~\cite{Bekenstein:1980jp,Casini:2008cr}, i.e.\ $S_{\rm exc} < 
\delta E/T_{\rm H}$ with $(\delta E/T_{\rm H} - S_{\rm exc})/S_{\rm exc} 
\nll 1$,%
\footnote{This assumption was also used implicitly in obtaining 
 Eq.~(\ref{eq:S-total}).}
and that a semiclassical excitation has entropy of order a few or larger, 
we obtain
\begin{equation}
  \mbox{a few} \lesssim S_{\rm exc} < \frac{\delta E}{T_{\rm H}} 
\quad\Rightarrow\quad
  \sin^2\!\theta \ll 1.
\label{eq:semicl}
\end{equation}
We thus find, unlike the claim in Ref.~\cite{Marolf:2015dia}, that 
a state having excitations over a semiclassical black hole background 
is atypical in the microscopic Hilbert space.

The direct application of the above analysis is limited to the 
excitations outside the horizon, i.e.\ the states obtained by 
acting $b_\gamma^\dagger$'s, which raise the energy of the 
state.  On the other hand, an object in the interior, excited 
by $a_\xi^\dagger$'s, involves operators $\tilde{b}_\gamma^\dagger$, 
which lower the energy as measured in the asymptotic region.  This 
is reflected in the fact that the infalling Hamiltonian $H$ in 
Eq.~(\ref{eq:H_eff}) does not commute with the generator of time 
evolution in the distant description, so that a positive energy 
excitation in the zone will develop negative energy components 
``after passing the horizon.''  However, the conclusion that 
a semiclassically excited state is atypical still persists if 
we focus only on excitations thrown from the exterior, i.e.\ 
the states that can be obtained by acting the infalling time 
evolution operator $U = e^{-i H \tau}$ on the states considered 
in the previous paragraph,%
\footnote{This excludes certain configurations that are obtained 
 by sending signals from the second exterior in a genuinely 
 two-sided black hole.  In the context of a collapse-formed 
 black hole, such configurations correspond to the states 
 obtained by performing highly fine-tuned and complicated 
 operations to the soft modes and/or radiation.}
since the operator $U$ is approximately unitary over the 
relevant timescale $\tau$.

This implies that the Hilbert space for semiclassical excitations, 
${\cal H}_{\rm exc}$, built on each of the orthogonal black hole and 
radiation microstates need not overlap significantly with each other. 
This is indeed expected to be the case from genericity consideration; 
we can show that states representing the same semiclassical excitation 
but built on different orthogonal microstates $A$ and $B$ have overlap
\begin{equation}
  {}_A \bra{\Psi(M)} {\cal O}_{\delta E}^{(A)\dagger} {\cal O}_{\delta E}^{(B)}
    \ket{\Psi(M)}_B 
  \approx O\biggl( \frac{1}{\sqrt{e^{S_{\rm bh}(M)} 
    e^{S_{\rm rad}}}} e^{-\frac{\delta E}{2 T_{\rm H}}} \biggr),
\label{eq:overlap}
\end{equation}
where ${\cal O}_{\delta E}^{(A,B)}$ are the operators that excite an appropriately 
smoothed semiclassical mode of energy $\delta E$ (either in a distant or 
infalling frame).  We find that this level of suppression is sufficient for 
us to be able to treat the microscopic Hilbert space as
\begin{equation}
  {\cal H} \approx {\cal H}_{\rm exc} \otimes {\cal H}_{\rm vac},
\label{eq:prod}
\end{equation}
where the elements of ${\cal H}_{\rm vac}$ cannot be discriminated as quantum 
degrees of freedom in the semiclassical theory.  Note that this structure 
is different from that considered in Refs.~\cite{Papadodimas:2012aq,%
Papadodimas:2013jku,Papadodimas:2015jra,Bousso:2013ifa,Marolf:2015dia}. 
In particular, we can define global operators ${\cal O} = b_\gamma, 
b_\gamma^\dagger, a_\xi, a_\xi^\dagger$ that act linearly throughout 
the space of all semiclassical states built on each of the vacuum microstates%
\footnote{These operators can be easily extended to linear operators 
 throughout the whole microscopic Hilbert space.  It is simply that they 
 then cannot be interpreted as annihilation and creation operators outside 
 the space of semiclassical states built on each vacuum microstate.}
\begin{equation}
  {\cal O} = \sum_{A = 1}^{e^{S_{\rm bh}(M) + S_{\rm rad}}} 
    \sum_{I} \sum_{J} \ket{\Psi_I^{(A)}} {\cal O}_{IJ}^{(A)} \bra{\Psi_J^{(A)}},
\label{eq:linear-op}
\end{equation}
where $I$ and $J$ are the indices specifying semiclassical states (regardless 
of the microstate), $A$ runs over orthogonal vacuum microstates of the 
form in Eq.~(\ref{eq:sys-state}), $\ket{\Psi_I^{(A)}}$ is the semiclassical 
state $I$ built on microstate $A$, and ${\cal O}_{IJ}^{(A)}$ is the matrix 
element of the corresponding operator built on $A$---$b_\gamma^{(A)}, 
b_\gamma^{(A)\dagger}, a_\xi^{(A)}, a_\xi^{(A)\dagger}$---in the microscopic 
Hilbert space.  These global operators, ${\cal O}$, obey the correct 
annihilation and creation operator algebra up to corrections exponentially 
suppressed in $\delta E/T_{\rm H}$, which is typically huge for a small 
object falling into the black hole.

Finally, we have divided the system into three components---the hard modes, 
soft modes, and radiation---rather than two.  This allows us to impose an 
energy constraint in the black hole system, which is essential in deriving 
the properties of operators in the effective theory described above. 
It also elucidates a difference between our construction and that of 
Refs.~\cite{Papadodimas:2012aq,Papadodimas:2013jku,Papadodimas:2015jra}; 
our effective second-exterior operators, $\tilde{b}_\gamma$ and 
$\tilde{b}_\gamma^\dagger$, are constructed as the mirror of hard 
modes, while those in Refs.~\cite{Papadodimas:2012aq,Papadodimas:2013jku,%
Papadodimas:2015jra} as the mirror of a system including early 
radiation.  This makes, for example, the vanishing of commutators 
between $\tilde{b}_\gamma$ / $\tilde{b}_\gamma^\dagger$ and 
operators acting on early radiation not automatic.

We may, however, expect that any simple operation performed on early 
radiation does not affect the black hole interior, protecting locality 
at the semiclassical level.  This possibility has recently been studied 
in Ref.~\cite{Kim:2020cds} using the concept of computational complexity. 
These authors have analyzed the problem for a state without an energy 
constraint (a state of the form in Eq.~(\ref{eq:sys-large}) in the next 
section); here we discuss the issue using a state, Eq.~(\ref{eq:sys-state}), 
that has a physically relevant form in our framework.  Consider the 
microscopic-level expression (see Eq.~(\ref{eq:coarse})) of a coarse-grained 
state $\ketc{\{ n_\alpha \}}$ built on microstate $A$
\begin{equation}
  \ketc{\{ n_\alpha \}}_A 
  = e^{\frac{E_n}{2 T_{\rm H}}} \sqrt{\sum_m e^{-\frac{E_m}{T_{\rm H}}}}\, 
    \sum_{i_n = 1}^{e^{S_{\rm bh}(M-E_n)}} \sum_{a = 1}^{e^{S_{\rm rad}}} 
    c_{n i_n a}^{(A)} \ket{\psi^{(n)}_{i_n}} \ket{\phi_a}.
\label{eq:coarse-A}
\end{equation}
A general operation on early radiation can be expressed as a quantum channel
\begin{equation}
  {\cal E}\left[\ketc{\{ n_\alpha \}}_A {}_A\brac{\{ n_\alpha \}}\right] 
  = \sum_x E_x \ketc{\{ n_\alpha \}}_A {}_A\brac{\{ n_\alpha \}} E_x^\dagger,
\label{eq:channel}
\end{equation}
where $E_x$ are Kraus operators that act on radiation as
\begin{equation}
  E_x = \sum_{a = 1}^{e^{S_{\rm rad}}} \sum_{b = 1}^{e^{S_{\rm rad}}} 
    \ket{\phi_a} E^x_{ab} \bra{\phi_b},
\label{eq:E_x}
\end{equation}
which satisfy $\sum_x E_x^\dagger E_x = \mathbb{I}$.

Let us now consider the matrix element of this state between coarse-grained 
states $\ketc{\{ \kappa_\alpha \}}$ and $\ketc{\{ \lambda_\alpha \}}$ built 
on microstate $B$ and $C$, respectively.  This gives
\begin{align}
  & {}_B\brac{\{ \kappa_\alpha \}}\, {\cal E}\left[\ketc{\{ n_\alpha \}}_A 
    {}_A\brac{\{ n_\alpha \}}\right]\, \ketc{\{ \lambda_\alpha \}}_C 
  = e^{\frac{2 E_n}{T_{\rm H}}} 
    \biggl( \sum_m e^{-\frac{E_m}{T_{\rm H}}} \biggr)^2 
    \delta_{\{ \kappa_\alpha \} \{ n_\alpha \}} 
    \delta_{\{ \lambda_\alpha \} \{ n_\alpha \}}
\nonumber\\
  & \qquad\qquad
  \times \sum_x \left[ \sum_{i_n = 1}^{e^{S_{\rm bh}(M-E_n)}} 
      \sum_{a = 1}^{e^{S_{\rm rad}}} \sum_{b = 1}^{e^{S_{\rm rad}}} 
      c_{n i_n a}^{(B)*} E^x_{ab} c_{n i_n b}^{(A)} 
    \right] 
    \left[ \sum_{j_n = 1}^{e^{S_{\rm bh}(M-E_n)}} 
      \sum_{c = 1}^{e^{S_{\rm rad}}} \sum_{d = 1}^{e^{S_{\rm rad}}} 
      c_{n j_n c}^{(A)*} E^x_{cd} c_{n j_n d}^{(C)} 
    \right].
\label{eq:after-operation}
\end{align}
We thus find that the condition for the coarse-grained states at the 
semiclassical level---and hence the interior---not to be affected by 
any measurement made by a remote observer is
\begin{equation}
  e^{\frac{E_n}{T_{\rm H}}} \biggl( \sum_m e^{-\frac{E_m}{T_{\rm H}}} \biggr) 
    \sum_{i_n = 1}^{e^{S_{\rm bh}(M-E_n)}} 
      \sum_{a = 1}^{e^{S_{\rm rad}}} \sum_{b = 1}^{e^{S_{\rm rad}}} 
      c_{n i_n a}^{(A)*} E^x_{ab} c_{n i_n b}^{(B)} 
  = q_x U^x_{AB}
\label{eq:pseudo-cond}
\end{equation}
for all $n = \{ n_\alpha \}$, $A$, $B$, and $x$, up to exponentially 
suppressed corrections.  Here, $U^x_{AB}$ are the elements of (arbitrary) 
unitary matrices $U^x$ acting on the space of microstates, and $q_x$ are 
numbers satisfying $\sum_x |q_x|^2 = 1$.  This is morally the condition 
for the simplicity of the measurement and pseudorandomness of Hawking 
radiation discussed in Ref.~\cite{Kim:2020cds}, which we expect to be 
satisfied under assumptions similar to those adopted there.  Assuming 
this is true, our framework preserves locality at the semiclassical 
level; i.e., any observer performing a simple operation on early radiation 
cannot remotely affect the black hole interior.%
\footnote{If an observer performs an operation with superpolynomial 
 complexity, he/she can affect the state of the black hole at the 
 semiclassical level; in fact, they can even create a firewall state. 
 Even in this case, however, the black hole quickly ``repairs'' itself 
 and recover the smooth horizon in a timescale of order the scrambling 
 time~\cite{Nomura:2018kia,Nomura:2019qps}.}

\section{Large AdS Black Hole}
\label{sec:v2}

After the original submission of this paper,%
\footnote{The original version of this paper was submitted such that 
 it would appear in an earlier announcement (submitted on Tue, 26 
 Nov 2019 14:02:29, EST).  The appearance, however, was delayed due 
 to a moderation by arXiv administration, which resulted in a larger 
 arXiv number.}
an interesting paper by Penington, Shenker, Stanford, and 
Yang appeared~\cite{Penington:2019kki} which discusses 
related issues.  In particular, these authors adopted the 
same machinery as that used here to construct interior operators 
in Eqs.~(\ref{eq:ann-av},~\ref{eq:cre-av}), which they referred 
to as the Petz map following the terminology used in quantum 
information science.

The models employed in Ref.~\cite{Penington:2019kki} are analogous 
to a large AdS black hole, which is distinct from a flat space or 
small AdS black hole.  For a large AdS black hole, there are modes 
whose wavelengths in the angular directions are smaller than the 
horizon radius and yet which have frequencies smaller than the 
Hawking temperature, since the inverse Hawking temperature is (much) 
smaller than the horizon radius, $1/T_{\rm H} \ll r_+$.  Here, 
$T_{\rm H} = 3 r_+/4\pi l^2$ is the Hawking temperature, and $r_+$ 
and $l$ are the horizon and AdS radii, respectively.  This is in 
contrast to the case of a flat space or small AdS black hole, in 
which $1/T_{\rm H} = 4\pi r_+$, so a mode having a wavelength in 
the angular directions smaller than the horizon radius necessarily 
has a frequency larger than $T_{\rm H}$.

Let us see what happens if we apply the construction in the previous 
section to these modes ($1/r_+ \ll \omega, \varDelta\omega \lesssim 
T_{\rm H}$), which we call ``relatively harder'' (but still soft) 
modes.  Since the uncertainty in energy ($\sim T_{\rm H}$) is larger 
than the frequencies of these modes, their states $\ket{\{ n_\alpha \}}$ 
need not be correlated with the states $\ket{\psi_i}$ of the other, 
``relatively softer'' (soft) modes ($\omega \lesssim 1/r_+$) as in 
Eq.~(\ref{eq:sys-state}).  Specifically, the state of the system with 
the black hole put in the semiclassical vacuum can be written as
\begin{equation}
  \ket{\Psi(M)} = \sum_{n = 1}^{e^{S_{\rm h}}} 
    \sum_{i = 1}^{e^{S_{\rm bh}(M)}} 
    \sum_{a = 1}^{e^{S_{\rm rad}}} c_{n i a} 
    \ket{\{ n_\alpha \}} \ket{\psi_i} \ket{\phi_a},
\label{eq:sys-large}
\end{equation}
where $e^{S_{\rm h}}$ ($\ll e^{S_{\rm bh}(M)}$) is the dimension 
of the Hilbert space for the harder modes, $S_{\rm bh}(M) = 
\pi (2 l^2 M/l_{\rm P})^{2/3}$ is the density of states for the 
softer modes ($\approx$ that of the black hole), and $\ket{\phi_a}$ 
represents states of the auxiliary system to which the AdS system 
is coupled at the boundary.%
\footnote{For simplicity, here we have assumed that the energy gaps 
 between different $\ket{\phi_a}$'s are smaller than of $O(T_{\rm H})$, 
 the precision with which the black hole mass is specified.  To see 
 the finiteness of the temperature in the auxiliary system (Hawking 
 radiation), we need to include states of the auxiliary system with 
 energy gaps larger than $T_{\rm H}$.}
Note that $\ket{\{ n_\alpha \}}$ and $\ket{\psi_i}$ here represent the 
states of the harder and softer soft modes, rather than hard and soft 
modes.  For simplicity, we set all the hard modes to be on their ground 
states; these modes will be considered later.

We can now define the normalized coarse-grained states along the 
lines of Eq.~(\ref{eq:coarse-prop}):
\begin{equation}
  \ketc{\{ n_\alpha \}} = \sqrt{e^{S_{\rm h}}} 
    \sum_{i = 1}^{e^{S_{\rm bh}(M)}} 
    \sum_{a = 1}^{e^{S_{\rm rad}}} 
    c_{n i a} \ket{\psi_i} \ket{\phi_a}.
\end{equation}
The statistical errors for the normalizations are fractionally of 
order $1/\sqrt{e^{S_{\rm bh}(M)} e^{S_{\rm rad}}} \sim 1/e^{S_{\rm sys}}$; 
here and below, we denote contributions of order $1/e^{\# S_{\rm bh} 
+ \#' S_{\rm rad}}$ simply by $1/e^{S_{\rm sys}}$.  We then find that 
the softer mode states that are entangled with different harder mode 
states have nonzero overlaps
\begin{equation}
  \innerc{\{ \kappa_\alpha \}}{\{ \lambda_\alpha \}} 
  = O\Biggl(\frac{1}{\sqrt{e^{S_{\rm bh}} e^{S_{\rm rad}}}}\Biggr) 
  \quad \mbox{for } \{ \kappa_\alpha \} \neq \{ \lambda_\alpha \},
\end{equation}
although they are small, of order $1/e^{S_{\rm sys}}$.  This is in 
contrast to the soft mode states entangled with the hard mode states 
in Sections~\ref{sec:framework} and \ref{sec:interior}, whose overlaps 
were virtually zero as in Eq.~(\ref{eq:soft-orthonorm}).

These small overlaps allow us, after the Page time, to choose 
``$\tilde{b}_\gamma$ and $\tilde{b}_\gamma^\dagger$ operators'' 
{\it for the harder soft modes} which act only on the auxiliary 
system, i.e.\ early radiation.  Consider operators
\begin{align}
  \tilde{b}_\gamma &= e^{S_{\rm h} + S_{\rm bh}} 
    \sum_{n = 1}^{e^{S_{\rm h}}} \sqrt{n_\gamma}\, 
    \sum_{i = 1}^{e^{S_{\rm bh}}} 
    \sum_{a = 1}^{e^{S_{\rm rad}}} \sum_{b = 1}^{e^{S_{\rm rad}}} 
    c_{n_- i a} c_{n i b}^*\, \ket{\phi_a} \bra{\phi_b},
\label{eq:ann-rad}\\
  \tilde{b}_\gamma^\dagger &= e^{S_{\rm h} + S_{\rm bh}} 
    \sum_{n = 1}^{e^{S_{\rm h}}} \sqrt{n_\gamma + 1}\, 
    \sum_{i = 1}^{e^{S_{\rm bh}}} 
    \sum_{a = 1}^{e^{S_{\rm rad}}} \sum_{b = 1}^{e^{S_{\rm rad}}} 
    c_{n_+ i a} c_{n i b}^*\, \ket{\phi_a} \bra{\phi_b}.
\label{eq:cre-rad}
\end{align}
This leads to
\begin{align}
  \brac{\{ \kappa_\alpha \}} \tilde{b}_\gamma 
    \ketc{\{ \lambda_\alpha \}} 
  &= \sqrt{\lambda_\gamma}\, 
    \delta_{\{ \kappa_\alpha \} \{ \lambda_\alpha - \delta_{\alpha\gamma} \}},
\\
  \brac{\{ \kappa_\alpha \}} \tilde{b}_\gamma^\dagger 
    \ketc{\{ \lambda_\alpha \}} 
  &= \sqrt{\lambda_\gamma + 1}\, 
    \delta_{\{ \kappa_\alpha \} \{ \lambda_\alpha + \delta_{\alpha\gamma} \}},
\end{align}
and
\begin{align}
  \brac{\{ \kappa_\alpha \}} \tilde{b}_\beta \tilde{b}_\gamma^\dagger 
    \ketc{\{ \lambda_\alpha \}} 
  &= \sqrt{\lambda_\beta + \delta_{\beta\gamma}} \sqrt{\lambda_\gamma + 1}\, 
    \delta_{\{ \kappa_\alpha \} 
      \{ \lambda_\alpha - \delta_{\alpha\beta} + \delta_{\alpha\gamma} \}}\, 
    \left[ 1 + O\Biggl(\frac{e^{S_{\rm bh}}}{e^{S_{\rm rad}}}\Biggr)\, 
      \delta_{\{ \kappa_\alpha \} \{ \lambda_\alpha \}} \right],
\label{eq:eff-rad-1}\\
  \brac{\{ \kappa_\alpha \}} \tilde{b}_\beta^\dagger \tilde{b}_\gamma 
    \ketc{\{ \lambda_\alpha \}} 
  &= \sqrt{\lambda_\beta - \delta_{\beta\gamma} + 1} \sqrt{\lambda_\gamma}\, 
    \delta_{\{ \kappa_\alpha \} 
      \{ \lambda_\alpha + \delta_{\alpha\beta} - \delta_{\alpha\gamma} \}}\, 
    \left[ 1 + O\Biggl(\frac{e^{S_{\rm bh}}}{e^{S_{\rm rad}}}\Biggr)\, 
      \delta_{\{ \kappa_\alpha \} \{ \lambda_\alpha \}} \right],
\label{eq:eff-rad-2}\\
  \brac{\{ \kappa_\alpha \}} \tilde{b}_\beta \tilde{b}_\gamma 
    \ketc{\{ \lambda_\alpha \}} 
  &= \sqrt{\lambda_\beta - \delta_{\beta\gamma}} \sqrt{\lambda_\gamma}\, 
    \delta_{\{ \kappa_\alpha \} 
      \{ \lambda_\alpha - \delta_{\alpha\beta} - \delta_{\alpha\gamma} \}},
\label{eq:eff-rad-3}\\
  \brac{\{ \kappa_\alpha \}} \tilde{b}_\beta^\dagger \tilde{b}_\gamma^\dagger 
    \ketc{\{ \lambda_\alpha \}} 
  &= \sqrt{\lambda_\beta + \delta_{\beta\gamma} + 1} \sqrt{\lambda_\gamma + 1}\, 
    \delta_{\{ \kappa_\alpha \} 
      \{ \lambda_\alpha + \delta_{\alpha\beta} + \delta_{\alpha\gamma} \}},
\label{eq:eff-rad-4}
\end{align}
where we have omitted contributions suppressed by $1/e^{S_{\rm sys}}$. 
In order for the operators in Eqs.~(\ref{eq:ann-rad},~\ref{eq:cre-rad}) 
to play the role of the annihilation and creation operators, the second 
terms in the square brackets in Eqs.~(\ref{eq:eff-rad-1},~\ref{eq:eff-rad-2}) 
must be negligible, which is the case if
\begin{equation}
  e^{S_{\rm bh}} \ll e^{S_{\rm rad}},
\label{eq:old-rad}
\end{equation}
i.e.\ the black hole is old.  With Eq.~(\ref{eq:old-rad}), 
the algebra of annihilation and creation operators, 
Eqs.~(\ref{eq:algebra-mat-1},~\ref{eq:algebra-mat-2}), 
is indeed satisfied.

The operators for the harder modes constructed in this way, however, are 
not relevant in erecting an effective theory of the interior.  This is 
because the majority of the states obtained by ``exciting'' a given vacuum 
microstate by these operators correspond simply to other vacuum microstates, 
i.e.\ states with different soft mode configurations.  One might think that 
there are some rare ``excitations'' generated by acting these operators 
on some microstate which can meaningfully be considered as being 
thrown into the black hole.  Such excitations, however, are quickly 
thermalized before hitting the stretched horizon.  In fact, they are 
nothing other than exponentially rare statistical fluctuations of the 
thermal soft mode gas, which do not represent a semiclassical object 
falling into the black hole.  This picture is consistent with the analysis 
of Ref.~\cite{Raju:2016vsu} in which it was shown that excitations having 
energy $\delta E \ll T_{\rm H}$ do not significantly affect relevant 
correlation functions in the bulk.

We therefore end up with the situation similar to the case of a 
flat space black hole.  A semiclassical object falling into the 
black hole is described by hard modes, which have frequencies 
$\omega$ and their gaps $\varDelta\omega$ larger than $\Delta$ 
determined by the black hole temperature
\begin{equation}
  \Delta \approx O\biggl(\frac{r_+}{l^2}\biggr).
\label{eq:Delta-large}
\end{equation}
Assuming that the energy of the black hole system (comprising the hard 
and soft modes) is specified with maximal precision of order $\Delta$, 
the state of the entire system is given by Eq.~(\ref{eq:sys-state}), 
where $\ket{\{ n_\alpha \}}$ and $\ket{\psi^{(n)}_{i_n}}$ represent 
the states of the hard modes and the corresponding (both harder and 
softer) soft modes, respectively.  We note, however, that the expression 
for the density of states now takes the form appropriate for a large 
AdS black hole, $S_{\rm bh}(M) = \pi (2 l^2 M/l_{\rm P})^{2/3}$.

As in the flat space case, the mass and entropy of the black hole can 
be viewed as being carried by the soft modes.  To see this, we note 
that the local temperature of these modes in the bulk is given by
\begin{equation}
  T_{\rm loc}(r) 
  = \frac{T_{\rm H}}{\sqrt{\frac{r^2}{l^2}-\frac{r_+^3}{l^2 r}}} 
  = \frac{3 r_+}{4\pi l^2 \sqrt{\frac{r^2}{l^2}-\frac{r_+^3}{l^2 r}}},
\label{eq:Tloc-large}
\end{equation}
and the stretched horizon is located at $r = r_{\rm s}$ with
\begin{equation}
  r_{\rm s} - r_+ \sim \frac{r_+ l_{\rm s}^2}{l^2}.
\label{eq:stretch-large}
\end{equation}
Here, $l_{\rm s}$ is the string length.  By integrating their entropy and 
energy densities, $\sim N T_{\rm loc}(r)^3$ and $\sim N T_{\rm loc}(r)^4$, 
from the stretched horizon toward the exterior, we find
\begin{alignat}{3}
  S &\sim N \int_{r_{\rm s}}^{\infty}\!\! T_{\rm loc}(r)^3 
    \frac{r^2 dr}{\sqrt{\frac{r^2}{l^2}-\frac{r_+^3}{l^2 r}}} 
  &&\sim \frac{r_+^2}{l_{\rm P}^2},
\label{eq:S-large}\\
  E_{\rm loc} &\sim N \int_{r_{\rm s}}^{\infty}\!\! T_{\rm loc}(r)^4 
    \frac{r^2 dr}{\sqrt{\frac{r^2}{l^2}-\frac{r_+^3}{l^2 r}}} 
  &&\sim \frac{M}{\sqrt{\frac{r_{\rm s}^2}{l^2}-\frac{r_+^3}{l^2 r_{\rm s}}}},
\label{eq:Eloc-large}
\end{alignat}
where $N$ is the number of low energy species, and we have used the 
relation $l_{\rm s}^2/N \sim l_{\rm P}^2$ and
\begin{equation}
  \frac{l}{\sqrt{r_+ (r_{\rm s}-r_+)}} 
  \sim \frac{1}{\sqrt{\frac{r_{\rm s}^2}{l^2}-\frac{r_+^3}{l^2 r_{\rm s}}}}.
\end{equation}
These indeed reproduce parametrically the entropy of the black hole 
and the mass $M$ measured at the stretched horizon, $E_{\rm loc}$, where 
most of the modes are located.  The internal dynamics of the soft modes 
near the stretched horizon is expected to be nonlocal below the AdS 
length scale $l$ in the directions along the horizon.

The construction of operators of an effective theory of the interior 
goes as in Section~\ref{sec:interior}.  Suppose that the state of the 
system with the black hole put in the semiclassical vacuum is given 
by Eq.~(\ref{eq:sys-state}) at time $t$.  The annihilation and creation 
operators for the infalling modes, $a_\xi$ and $a_\xi^\dagger$ in 
Eqs.~(\ref{eq:a_xi},~\ref{eq:a_xi-dag}), can then be constructed by 
$b_\gamma$ and $b_\gamma^\dagger$ in Eqs.~(\ref{eq:ann},~\ref{eq:cre}) 
and $\tilde{b}_\gamma$ and $\tilde{b}_\gamma^\dagger$ in 
Eqs.~(\ref{eq:ann-m-orig},~\ref{eq:cre-m-orig}).  If the black 
hole is young, i.e. if it is not maximally entangled with the 
rest of the system, then the operators $\tilde{b}_\gamma$ and 
$\tilde{b}_\gamma^\dagger$ can be taken to act only on the soft modes 
as in Eqs.~(\ref{eq:ann-av},~\ref{eq:cre-av}) with Eq.~(\ref{eq:c}). 
This option, however, is not available if the black hole is old. 
The erected effective theory describes the physics in the causal 
domain of the equal-time hypersurface at $t$ in the emergent 
effective two-sided black hole geometry.

Finally, we emphasize that the discussion at the end of Section~\ref{sec:interior} 
leading to Eq.~(\ref{eq:prod}) carries over to the case of a large AdS 
black hole.  Specifically, the overlap, $\sin\theta$, between a state in 
which hard modes are excited by $b_\gamma^\dagger$'s and a typical state 
of the same energy (within uncertainty of order $T_{\rm H} \sim r_+/l^2$) 
is given by
\begin{equation}
  \sin^2\!\theta \sim 
    e^{-\left( \frac{\delta E}{T_{\rm H}} - S_{\rm hard} \right)},
\label{eq:atypical-large}
\end{equation}
where $\delta E$ and $S_{\rm hard}$ are the energy and coarse-grained 
entropy of the hard mode excitation, respectively.  Therefore, assuming 
that the hard mode excitation is well within the Bekenstein bound, 
we obtain
\begin{equation}
  \mbox{a few} \lesssim S_{\rm hard} < \frac{\delta E}{T_{\rm H}} 
\quad\Rightarrow\quad
  \sin^2\!\theta \ll 1.
\label{eq:semicl-large}
\end{equation}
Given that the infalling time evolution operator $U = e^{-i H \tau}$ 
is approximately unitary over the relevant timescale, this translates 
into the statement that the states which have semiclassical objects 
falling inside the horizon occupy only a negligible fraction of the 
microscopic Hilbert space.  As discussed in Section~\ref{sec:interior}, 
this allows us to treat the microscopic Hilbert space as
\begin{equation}
  {\cal H} \approx {\cal H}_{\rm exc} \otimes {\cal H}_{\rm vac},
\label{eq:prod-large}
\end{equation}
where ${\cal H}_{\rm exc}$ and ${\cal H}_{\rm vac}$ are the Hilbert 
space of the effective semiclassical theory and that for the vacuum 
microstates, respectively.  In particular, physically relevant operators 
in the semiclassical theory---$b_\gamma$, $b_\gamma^\dagger$, $a_\xi$, 
and $a_\xi^\dagger$---can be globally defined as linear operators 
at the microscopic level, which obey the correct annihilation and 
creation operator algebra in the space of the semiclassical states 
up to corrections suppressed by $e^{-\delta E/T_{\rm H}}$.

\subsubsection*{Relation to entanglement wedge reconstruction and islands}

Recently, there has been significant progress in understanding 
the interior of a black hole using holographic entanglement wedge 
reconstruction~\cite{Penington:2019npb,Almheiri:2019psf,Almheiri:2019hni}. 
According to these analyses, operators acting on early radiation are 
sufficient to reconstruct a portion of the black hole interior after 
the Page time.  On the other hand, we have seen that in order to erect 
the effective theory of the interior at any given time in a distant 
description, the interior operators must act on the soft mode degrees 
of freedom in addition to the early radiation.  How can these two 
statements be reconciled?

A key to understand this issue is time evolution.  In general, 
entanglement wedge reconstruction assumes that we know the time 
evolution operator of the holographic theory, i.e.\ that of a system 
comprising the boundary theory and any auxiliary theory coupling to it. 
In addition, the analyses referred to above assume that the information 
leaked from the boundary theory---representing the bulk with a black 
hole---to the auxiliary system---a system storing Hawking radiation---is 
effectively irreversible.  These imply that given a state at some time 
$t$, we can adopt the following strategy to reconstruct the interior. 
Since a hard mode object that had fallen before some time $t_{\rm w}$ 
($< t$), as well as the soft modes entangled with it, would be fully 
mixed into the black hole, and their information is emitted in radiation 
through time evolution, we can in principle represent (a portion of) 
the interior spacetime describing the fate of the object on radiation 
at time $t$.  If we have a complete knowledge about the radiation 
state at $t$, then $t - t_{\rm w}$ is of the order of the information 
retention time, but if we lack a part of the knowledge, then 
$t - t_{\rm w}$ can be larger.

This explains why the interior portion of the entanglement wedge of 
radiation in the work of Refs.~\cite{Penington:2019npb,Almheiri:2019psf,%
Almheiri:2019hni} emerges as an island:\ the region disconnected from 
that supporting the radiation.  We can indeed check the consistency of 
this understanding at the level of precision including the coefficients 
of terms enhanced by the logarithm of the black hole mass or entropy.

\begin{figure}[t]
\begin{center}
  \includegraphics[height=8cm]{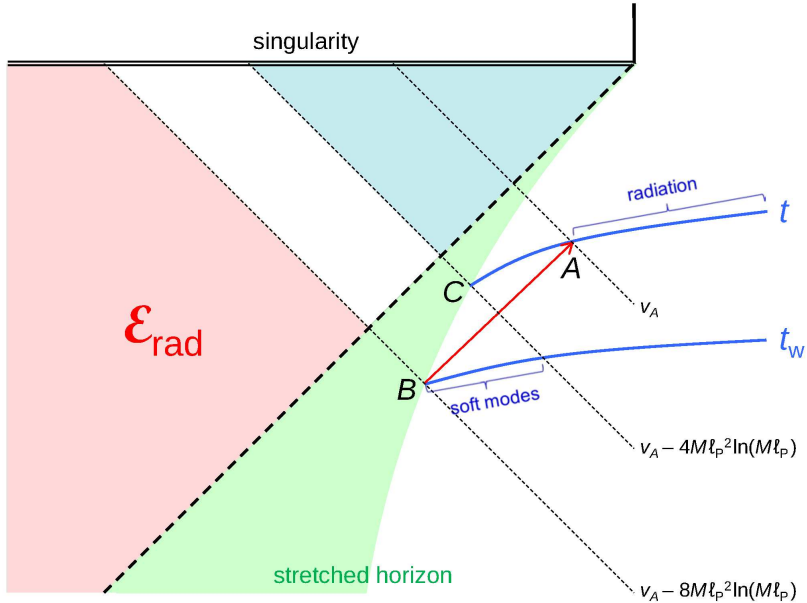}
\end{center}
\caption{Radiation of a state given at time $t$ can represent the 
 hard modes hitting the stretched horizon at or before $t_{\rm w}$ 
 ($< t$) and the soft modes entangled with them, through time 
 evolution.  This allows us to construct the effective interior 
 theory erected at $t_{\rm w}$ whose operators act only on radiation 
 at $t$; these operators, however, cannot describe the future of the 
 zone at $t_{\rm w}$, since the radiation cannot represent hard modes 
 there nor the soft modes entangled with them.  A similar construction 
 works for effective theories erected at times earlier than $t_{\rm w}$, 
 implying that we can reconstruct the spacetime region denoted 
 by ${\cal E}_{\rm rad}$, which reproduces the entanglement wedge 
 of the radiation at time $t$.  This region, however, cannot 
 describe the fate of a falling object located in the zone at 
 time $t$ (segment~$CA$), which occurs in the future of $C$ 
 (shaded).  To construct an effective theory describing (a part 
 of) this region building on the state at time $t$, we need operators 
 that act both on the soft modes and radiation (in addition to those 
 acting on the hard modes).}
\label{fig:wedge}
\end{figure}

For concreteness, let us consider a black hole in 4-dimensional asymptotically 
flat spacetime.  Suppose that the state is given at a boundary ($\approx$ 
Schwarzschild) time $t$ after the Page time.  The state of the 
radiation then represents the $r \gtrsim r_{\rm z}$ portion of this 
equal-time hypersurface, where $r_{\rm z}$ is the location of the 
edge of the zone, and we denote the spacetime point $(t,r_{\rm z})$ 
by $A$; see Fig.~\ref{fig:wedge}.
Now, given this radiation state at time $t$, we can determine the 
state of hard mode excitations that had fallen into the stretched 
horizon before $t_{\rm w}$ if $t - t_{\rm w}$ is of the order of 
the scrambling time or larger~\cite{Hayden:2007cs,Sekino:2008he}. 
(The required $t - t_{\rm w}$ becomes larger if the amount of 
information carried by the excitations is large.)  Since the black 
hole information is contained in the soft modes, however, we expect 
that this time, $t - t_{\rm w}$, is larger than the signal propagation 
time between the $r = r_{\rm s}$ (point~$B$ in Fig.~\ref{fig:wedge}) 
and the edge of the zone $r = r_{\rm z}$ (point~$A$):
\begin{equation}
  t - t_{\rm w} \gtrsim 4Ml_{\rm P}^2 \ln(M l_{\rm P}) + O(Ml_{\rm P}^2).
\label{eq:t-tw}
\end{equation}
Note that the radiation state at $t$ cannot determine a hard mode 
excitation at time $t_{\rm w}$ unless it is hitting the stretched 
horizon, since the relevant hard modes will not be scrambled by the 
time $t$.  In other words, hard mode operators away from the stretched 
horizon cannot be represented on the radiation at time $t$.

The fact that hard modes falling before $t_{\rm w}$ are fully scrambled 
in the black hole allows us to avoid the energy constraint of the form 
of Eq.~(\ref{eq:sys-state}) for these modes, as well as the modes entangled 
with them, and hence to construct corresponding operators of the effective 
theory at time $t_{\rm w}$ acting on the state of radiation at time $t$. 
These operators, however, cannot describe the future of the zone at $t_{\rm w}$ 
because hard modes there are not scrambled.  A similar construction also 
works for effective theories before $t_{\rm w}$.  This implies that if the 
scrambling time is as small as the signal propagation time, then we can 
reconstruct bulk operators in the region denoted by ${\cal E}_{\rm rad}$ 
in the figure.  In $3+1$ dimensions, this indeed reproduces the entanglement 
wedge of the radiation system given in Ref.~\cite{Penington:2019npb}.%
\footnote{For simplicity, here we have ignored the stretching inside 
 the horizon because it is not relevant for our discussion.}
In $d+1$ dimensions ($d > 3$), the scrambling and signal propagation 
times are different, and $t - t_{\rm w}$ must be taken as the scrambling 
time.  We can, however, still check that the scrambling time obtained 
in Ref.~\cite{Penington:2019npb} is indeed larger, by a factor of $(d-1)/2$, 
than the corresponding signal propagation time between the stretched 
horizon and the location where Hawking radiation is extracted.

In our view, entanglement wedge reconstruction of the interior using 
only radiation degrees of freedom describes a collection of (a portion 
of) spacetime regions associated with effective interior theories 
that {\it could be erected} in the past, specifically at times earlier 
than the scrambling time before $t$.  This leads to some issues 
(though not necessarily problems) from the viewpoint of actually 
constructing operators describing the interior.  First, since the 
reconstruction involves time evolution backward in time, the expressions 
for the bulk operators in terms of boundary operators are highly 
complicated, and the reconstructed operators are extremely fragile; 
i.e., a small deformation of boundary operators destroys the success 
of the reconstruction.  More importantly, the reconstruction does 
not provide operators in the interior region that are relevant for 
describing the fate of an object that is located in the zone region 
at the time when the state is given.

This can be seen in Fig.~\ref{fig:wedge}.  In terms of the ingoing 
Eddington-Finkelstein coordinate $v = t + r_*$, where $r^* = r + 
2Ml_{\rm P}^2 \ln[(r-2Ml_{\rm P}^2)/2Ml_{\rm P}^2]$ is the tortoise 
coordinate, point~$B$ is $8Ml_{\rm P}^2 \ln(M l_{\rm P})$ earlier than 
point~$A$.  Namely, the entanglement wedge of the radiation (at time 
$t$) describes only the
\begin{equation}
  v < v_B = v_A - 8Ml_{\rm P}^2 \ln(M l_{\rm P}) + O(Ml_{\rm P}^2)
\label{eq:region-1}
\end{equation}
portion of the interior spacetime.  On the other hand, to describe 
the future of a falling object that is in the zone at time $t$ 
(segment~$CA$ in the figure), we need the portion
\begin{equation}
  v > v_C = v_A - 4Ml_{\rm P}^2 \ln(M l_{\rm P}) + O(Ml_{\rm P}^2),
\label{eq:region-2}
\end{equation}
and the two regions in Eqs.~(\ref{eq:region-1}) and 
(\ref{eq:region-2}) do not overlap.  In order to erect an effective 
theory that is capable of describing future evolution of such an 
object, we need to use operators that act both on the soft modes and 
radiation (or, equivalently, on radiation at a sufficiently later time), 
as discussed in this paper.

\section{Conclusions and Discussion}
\label{sec:concl}

In this paper, we have shown that operators describing the experience 
of an observer falling into the horizon can be constructed without 
contradicting the unitary evolution of the black hole.  The choice 
of these operators at the microscopic level is not unique.  In particular, 
for a young black hole, we can choose them to act only on the degrees 
of freedom that are directly associated with the black hole:\ the hard 
and soft modes.  On the other hand, for an old black hole, the operators 
must also act on radiation emitted earlier.  The difference between the 
two cases comes from the statistics associated with the coarse-graining 
performed to obtain the effective theory of the interior.  We 
have also discussed relation between the present construction 
and entanglement wedge reconstruction of the interior described 
in Refs.~\cite{Penington:2019npb,Almheiri:2019psf,Almheiri:2019hni}.

Before concluding, let us discuss the origin of the coarse-graining 
from a slightly different perspective.  From general considerations, 
we know that the dimension of the Hilbert space describing semiclassical 
physics in the interior region, $e^{S_{\rm int}}$, is much smaller 
than the number of independent black hole or radiation microstates:\ 
$S_{\rm int} \ll S_{\rm bh} \sim S_{\rm rad}$ (except possibly at the 
very beginning and end of the black hole evolution).  This implies 
that to erect an effective theory of the interior, we must find 
very special degrees of freedom within those of the black hole 
and/or radiation which are relevant for describing the semiclassical 
physics in the interior spacetime.  How can such degrees of 
freedom be selected?

One way to identify these degrees of freedom is to utilize a subset 
of semiclassical modes in the exterior of the horizon.  The relevant 
degrees of freedom are then those entangled with these exterior 
modes, as they allow us to construct operators satisfying the 
correct annihilation and creation operator algebra up to errors 
of order $e^{-S_{\rm sys}}$.  This, however, does not necessarily 
lead to the picture of semiclassical interior spacetime.  In particular, 
for a large AdS black hole we can apply this procedure to exterior 
modes whose wavelengths are smaller than the horizon size, finding 
operators that satisfy the correct algebra for each microstate. 
These operators, however, cannot be extended to linear operators 
defined throughout the microstates if the frequencies of the modes 
are smaller than the Hawking temperature.  In other words, the Fock 
spaces built by acting these operators on each microstate significantly 
overlap with each other.

At first sight, this seems to force us to embrace ``state-dependence'' 
of the interior operators in the sense of Refs.~\cite{Papadodimas:2012aq,%
Papadodimas:2013jku,Papadodimas:2015jra}, since it means that the 
same operator has multiple interpretations.  This is, however, not 
the case if we adopt the view, as we did in this paper, that in a 
distant description the black hole microstates are represented by 
the configurations of the soft modes.  In this case, ``exciting'' 
a given vacuum microstate by these operators corresponds simply to 
obtaining other vacuum microstates; a ``creation operator'' viewed 
from one vacuum microstate can be viewed as an ``annihilation operator'' 
(or a superposition of annihilation and creation operators) from 
the viewpoint of another vacuum microstate.

By definition, the physics at the semiclassical level should not 
depend sensitively on the microstate of the black hole or early 
radiation.  This implies that the operators discussed above are not 
relevant in constructing the theory of the interior at the semiclassical 
level.  The relevant operators are those that act on the modes 
having frequencies larger than the Hawking temperature---the hard 
modes---and the degrees of freedom entangled with them.  We have 
shown that the Fock spaces built by these operators on each of the 
orthogonal vacuum microstates have overlaps exponentially suppressed 
in the ratio of the energy of semiclassical excitation, $E_{\rm exc}$, 
to the Hawking temperature, $T_{\rm H}$.  We can therefore define these 
operators consistently throughout the microstates up to corrections 
of $e^{-O(E_{\rm exc}/T_{\rm H})}$.  These corrections should be 
viewed as an intrinsic ambiguity of the semiclassical theory.

We conjecture that the emergence of semiclassical physics, in fact, 
requires the existence of operators that can be applied globally 
on a set of microstates with (approximately) invariant meaning, 
as we have found here.  This allows us to view the microscopic 
Hilbert space as a direct product of the form ${\cal H}_{\rm exc} 
\otimes {\cal H}_{\rm vac}$ as far as these operators---or observables 
constructed out of these operators---are concerned.  Semiclassical 
theories are those describing the physics associated with the 
${\cal H}_{\rm exc}$ factor, which is insensitive to the microscopic 
physics occurring in the ${\cal H}_{\rm vac}$ part.

We think this is one of the main lessons we have learned from the 
paradoxes raised regarding the interior of an evaporating black hole. 
It is our hope that the picture presented in this paper sheds light 
on how quantum gravity works at the most fundamental level.

\section*{Acknowledgments}

I would like to thank the organizers and participants of the KITP 
conference ``Geometry from the Quantum''; lively and enjoyable 
discussions there have motivated me to make a revision integrating 
the contents of Section~\ref{sec:v2} into the main part in a 
cohesive manner.  This work was supported in part by the Department 
of Energy, Office of Science, Office of High Energy Physics under 
contract DE-AC02-05CH11231 and award DE-SC0019380, by MEXT KAKENHI 
Grant Number 15H05895, and by World Premier International Research 
Center Initiative (WPI Initiative), MEXT, Japan.

\end{document}